\documentclass[%
  aps,
  prl,
  superscriptaddress, letterpaper,
  twocolumn,
  amsfonts, amssymb, amsmath]{revtex4-2}
  
\usepackage[utf8]{inputenc}
\usepackage{amsmath}
\usepackage{amsfonts}
\usepackage{amssymb}
\usepackage{graphicx,color}

\begin{document}

\title{Lagrangian dispersion in experimental stratified turbulence}
\author{Maelys Magnier}
\affiliation{Laboratoire des Ecoulements Geophysiques et Industriels (LEGI), Universite Grenoble Alpes, CNRS, Grenoble-INP,  F-38000 Grenoble, France}
\author{Costanza Rodda}
\affiliation{Laboratoire des Ecoulements Geophysiques et Industriels (LEGI), Universite Grenoble Alpes, CNRS, Grenoble-INP,  F-38000 Grenoble, France}
\affiliation{Department of Civil and Environmental Engineering, Imperial College London, SW7 2AZ, UK}
\author{Cl{\'e}ment Savaro}
\affiliation{Laboratoire des Ecoulements Geophysiques et Industriels (LEGI), Universite Grenoble Alpes, CNRS, Grenoble-INP,  F-38000 Grenoble, France}
\affiliation{Laboratoire de {M\'ecanique} et ses Interfaces, ENSTA Paris, Institut Polytechnique de Paris, F-91120 Palaiseau, France}
\author{Pierre Augier}
\author{Nathanael Machicoane}
\author{Thomas Valran}
\author{Samuel Viboud}
\author{Nicolas Mordant}
\affiliation{Laboratoire des Ecoulements Geophysiques et Industriels (LEGI), Universite Grenoble Alpes, CNRS, Grenoble-INP,  F-38000 Grenoble, France}
\email[]{nicolas.mordant@univ-grenoble-alpes.fr}

\begin{abstract}
Lagrangian measurements of tracer particle dispersion in stratified turbulence are presented from a large-scale experiment achieving both high buoyancy Reynolds numbers and low Froude numbers -- a regime characteristic of oceanic conditions. Stratification has a pronounced effect on the vertical particle dispersion, which is observed to be constrained to distances on the order of the buoyancy scale $w_{\mathrm{std}}/N$, where $w_{\mathrm{std}}$ is the standard deviation of the vertical velocity and $N$ is the Brunt-Väisälä frequency. 
As expected in strongly nonlinear, stratified turbulence, the frequency spectrum of the Lagrangian velocity becomes isotropic at frequencies higher than $N$. The spectral decay follows a $1/f^3$ scaling, which contrasts with the $1/f^2$ behavior typical of homogeneous isotropic turbulence. 
At time scales corresponding to internal waves, the statistics of velocity increments remain Gaussian, consistent with the weakly nonlinear regime of wave turbulence. At smaller scales, however, the flow exhibits strongly non-Gaussian statistics, indicative of fully nonlinear turbulent dynamics driven by wave breaking. 
\end{abstract}


\maketitle

This Letter presents an experimental investigation of Lagrangian transport in stably stratified turbulence -- that is, turbulence occurring in a fluid with a stable vertical density gradient. Density stratification is a common feature in atmospheric, oceanic, and astrophysical flows \cite{Vallis,aerts_asteroseismology_2010}. In the ocean, it arises from variations in temperature or in the concentration of components such as salt or humidity. The presence of buoyancy forces associated with the density gradient significantly influences the turbulent dynamics, introducing anisotropy due to variations in potential energy in addition to kinetic energy. Thus, stratification has a major impact on the vertical fluxes of heat, carbon or other concentrations, which are key parameters to infer the role of the ocean in the impact of climate change \cite{Li2020,cheng_2023}.

Such turbulence is characterized by the Reynolds number $Re=uL_h/\nu$ (here based on the typical horizontal velocity $u$ and horizontal length scale $L_h$, $\nu$ is the kinematic viscosity), which should be very large and the horizontal Froude number $F_h=u/L_hN$, which should be much smaller than 1 for the flow to be dominated by gravity effects. $N$ is the Brunt Väisälä frequency: $N=\sqrt{-g(d\rho/dz)/\rho_0}$ ($g$ is the acceleration of gravity, $\rho(z)$ is the vertical density profile at rest and $\rho_0$ is the average density). In stratified flows, internal gravity waves can propagate with the angular frequency $\omega=N\sin \theta$ with $\theta$ the angle of the wavevector to the vertical~\cite{Lighthill}. At large enough Reynolds number, or more precisely at large enough buoyancy Reynolds number $Re_b=ReF_h^2\gtrsim 1$, waves are coupled by nonlinearity, yielding a regime of wave turbulence \cite{nazarenko2011wave,caillol2000kinetic,Labarre2024,cortet2025} . For $Re_b\gg 1$ strongly non linear stratified turbulence (SNLST) can develop \cite{brethouwer2007}; a range of scales is associated with strongly nonlinear anisotropic motion induced, for example, by wave breaking. Nonlinearity impacts the stratification through irreversible diffusion of temperature or irreversible mixing of components, with major impacts on the global ocean circulations, for example~\cite{mackinnon2017climate,caulfield2021}.

Transport and mixing in fluid flows are fundamentally linked to the Lagrangian motion of fluid particles \cite{Taylor,sawford2001,yeung2002,toschi2009,salazar2009}. A central question in transport theory is how far a fluid particle is displaced from its initial position by the flow. Similarly, mixing is concerned with how the distance between two fluid particles evolves over time -- either how initially close particles separate or how distant particles approach each other. The foundations of Lagrangian transport in turbulent flows were laid by Taylor in his seminal work~\cite{Taylor}.
In homogeneous stationary turbulence, the dispersion of a single fluid particle $\Delta_i^2(t)=\langle (x_i(t)-x_i(0))^2\rangle$ follows two asymptotic behaviors. For $t\ll T^i_L=\int_0^{+\infty}\langle u_i(t+\tau)u_i(t)\rangle d\tau/{u_i}_{std}^2$, one observes a ballistic regime $\Delta_i(t)={u_i}_{std}t$ (with ${u_i}_{std}$ the standard deviation of the $i$-th component of the velocity). At large times ($t\gg T_L^i$), a diffusive regime is observed with $\Delta_i^2(t)=2{u_i}_{std}^2T^i_L t$. These behaviors are fundamentally related to the correlations of the Lagrangian velocity $C_i(\tau)=\langle u_i(t+\tau)u_i(t)\rangle$ that have been studied numerically and experimentally in homogeneous isotropic turbulence \cite{Vanoni1955,Frenzen1963,yeung1989,mordant2001,toschi2009}. However, plume analysis \cite{britter1983} or numerical simulations of Lagrangian trajectories \cite{kimura1996,kaneda2000,cambon2004,liechtenstein2006,venayagamoorthy2006,aartrijk2008,brethouwer2009, sujovolsky2018,buaria2020,petropoulos2025} show that, in stratified turbulence, the vertical dispersion is significantly altered by gravity and the ballistic regime is rather followed by $\Delta_z(t)=constant$, possibly followed at much later time by a diffusive regime associated with molecular diffusion. The horizontal dispersion remains qualitatively similar to the case of isotropic turbulence. 
To the best of our knowledge only the pioneering work of Frenzen \cite{Frenzen1963} reports laboratory Lagrangian investigations in stratified turbulence that confirm this behavior. Frenzen's experiments revealed changes in velocity correlations between homogeneous and stratified fluids. However, the limited duration and number of measured particle trajectories in that study significantly constrain the scope of the observations.

The experimental setup closely follows that described in \cite{Rodda2022,Rodda2024}. Experiments are conducted in the Coriolis facility at LEGI, a circular tank with a diameter of 13~m, filled with water to a depth of $H = 1$m. Such a large facility is required to achieve both a large Reynolds number and a small Froude number. The fluid is stably density-stratified by varying salt and alcohol concentrations, producing a linear density profile with an initially uniform Brunt-Väisälä frequency of approximately $N = 0.25$ rad/s, while ensuring optical index matching. The flow is forced within a pentagonal region of the tank, bounded by vertical walls of horizontal length 6 m and extending over the entire fluid depth (see the Supplemental Material (SM) \cite{SM}). A pentagonal geometry was chosen to avoid having parallel side walls that would allow simple standing waves to occur by bouncing back and forth between the walls. Four of the walls can oscillate around a horizontal shaft located at mid-depth, thereby generating four broad beams of vertical mode-1 internal gravity waves. Each wavemaker operates at a frequency set to $0.7N$, with a slight random temporal modulation to prevent a possible strong resonance of a wave mode of the pentagon at the forcing frequency. The panels operate simultaneously with independent modulations. This configuration injects energy through large-scale waves, in analogy with oceanic submesoscale processes forced by large-scale tidal oscillations \cite{mackinnon2017climate}. As reported in \cite{Rodda2024}, when the value of $Re_b$ exceeds about 10, wave breaking is observed (see fig.~2 of \cite{Rodda2024}), so that in the conditions reported here, wave breaking occurs in all experiments. The fluid is seeded with solid polystyrene particles of diameter 700~$\mu$m, whose density is adjusted to be neutrally buoyant at mid-depth, within a resting layer of thickness about 0.3 m. Since their size is smaller than the Kolmogorov scale, estimated to be of the order of one millimeter, the particles are expected to behave as passive tracers. The central region of the pentagon is illuminated by eight underwater LED projectors (100 W each), while four PCO Edge 5.5 cameras (5.5 Mpixel sCMOS sensor, 16 bits) observe a volume around the flow center. The optical system is calibrated following the procedure of \cite{Machicoane2019}, yielding a measurement volume of $80 \times 80 \times 70$ cm$^3$ at the tank center in the common field of view of all four cameras (see SM \cite{SM}). Image acquisition is performed at 20 frames/s. Particle trajectories are reconstructed in time and space using the PTV algorithm described in \cite{PTV,Bourgoin2020}, with examples provided in the SM \cite{SM}. Only particle trajectories lasting more than 5~s are retained for analysis. Track durations are broadly distributed, with an average of 18 s and maximum values exceeding 200~s. About 250 such tracks are measured simultaneously. A typical experimental run begins with a 20-minute forcing phase, during which the wavemakers are operated to allow waves to propagate throughout the tank and turbulence to develop. Video recording is then performed for 90 minutes with a total of $108,000$ images for each camera (total 4.7 TB of uncompressed raw images for each experiment). The flow is not strictly stationary, as mixing progressively develops during the experiment, leading to the slow growth of mixed layers at the top and bottom of the tank. In the most strongly forced case, these layers reached about 15 cm in thickness, but remained outside the observed volume. Four experiments were conducted with increasing forcing amplitude, exploring regimes of stratified turbulence characterized by a large Reynolds number $Re > 10^4$, a small horizontal Froude number $F_h < 0.05$, and a buoyancy Reynolds number $Re_b = Re F_h^2$ significantly greater than unity (see Table~\ref{table_exps}). Eulerian measurements under similar conditions have been reported in \cite{Rodda2024}; the present experiments achieve higher $Re_b$. These studies show that, at large $Re_b$, the flow remains strongly anisotropic and exhibits similarities with SNLST \cite{brethouwer2007}.

\begin{table}[htb]
\begin{ruledtabular}
{\begin{tabular}{lcccccccccc}
 &$A$ (cm) & $Re$&$F_h$&$Re_b$ \\
\colrule
EXP1 & 7.5 &$2.9 \times 10^4$ & $2.6  \times 10^{-3}$ &19\\
EXP2& 9  & $3.5 \times 10^4$ & $3.1  \times 10^{-2}$ &33\\
EXP3 & 11  & $3.9 \times 10^4$ & $3.8  \times 10^{-2}$ &56\\
EXP4 & 13  & $4.6 \times 10^4$ & $4.5  \times 10^{-2}$ & 92
\end{tabular}}
\end{ruledtabular}
\caption{\label{table_exps}Parameters for the experimental runs. $A$ is the oscillation amplitude of the wavemaker, and $F=0.7$ is the forcing frequency normalized by $N$. We define the following nondimensional numbers with the forcing parameters: $Re=\frac{2ANH}{\nu}\sqrt{1-F^2}$ is the Reynolds number, $F_h=\frac{A}{2H}\frac{F^2}{\sqrt{1-F^2}}$ is the horizontal Froude number and $Re_b=ReF_h^2$ is the buoyancy Reynolds number \cite{Rodda2022}.}
\end{table}

\begin{figure}[!htb]
\includegraphics[width=8.5cm]{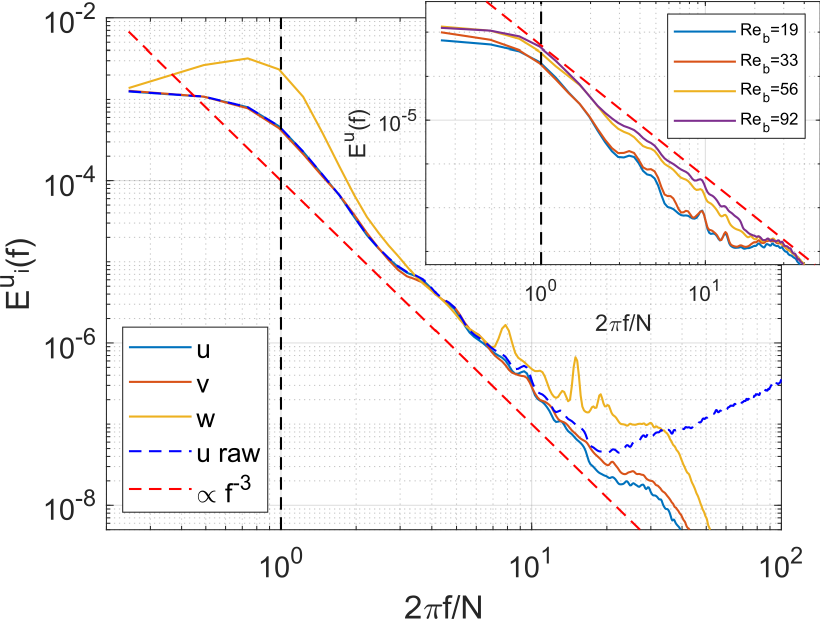}
\caption{\label{spvit} Lagrangian velocity power spectra. Fourier transforms are computed only on trajectories longer than 1024 frames. Main figure: EXP4. The dashed blue line shows the raw velocity $u$ spectrum, while the solid lines spectra are computed from the low-pass filtered velocity components, removing the noise-dominated high-frequency part. The vertical dashed line highlights $f=N/2\pi$. The red dahsed line is a $f^{-3}$ decay. Insert: Spectra of $u$ for all experiments.}
\end{figure}

The power spectral density of the velocity of the particles is shown in fig.~\ref{spvit} for EXP4. The duration of the tracks is not long enough to properly resolve the wave dominated range of frequencies ($f<N/2\pi$) but the few available points show that the flow is anisotropic at these frequencies, as observed in the Eulerian frequency spectrum, due to a kinematic relation between the vertical and horizontal components of internal gravity waves~\cite{savaro2020generation}. When increasing the frequency beyond $N/2\pi$ (out of the wave range), the spectra become isotropic for $f>4N/2\pi$ with the spectrum of the vertical component being very close to that of the horizontal components. The separation of the curves at frequencies $2\pi f/N\gtrsim10$ comes from experimental noise that is stronger on the vertical component of the velocity. Note that, for the analysis, the trajectories have been lowpass filtered by a Gaussian kernel as in \cite{mordant2004}. The blue dashed line displays the raw data spectrum and shows that this data is dominated by noise for $2\pi f/N>20$. In the  frequency range $2<2\pi f/N<20$ a clear power law decay is observed as $\propto 1/f^3$. This $-3$ exponent is steeper than the $-2$ exponent observed in Homogeneous Isotropic Turbulence (HIT) \cite{yeung1989,mordant2001} and predicted by dimensional analysis in the spirit of 1941 Kolmogorov's theory. Eulerian spectra (shown in SM \cite{SM}) in conditions similar to that of EXP4 show a decay close to $1/f^{2.5}$, so the Lagrangian spectrum is decaying faster than its Eulerian counterpart. This observation is qualitatively similar to that of HIT for which the Eulerian spectrum decays as $1/f^{5/3}$ and the Lagrangian one as $1/f^2$ \cite{Chevillard}. This difference is due to the occurrence of sweeping of small structures by large-scale eddies \cite{Tennekes_1975}. Sweeping has been observed in similar conditions of high $Re_b$ in \cite{Rodda2024} (see their fig.~4 and associated discussion), thus a similar effect may be occurring also in strongly nonlinear stratified turbulence. The insert of fig.~\ref{spvit} shows that the decay of the spectrum is steeper at the lowest values of $Re_b$ (EXP1 and EXP2) and that it goes to $1/f^3$ when increasing $Re_b$, i.e. when large scale waves keep breaking as described in \cite{Rodda2024}. This observations share some similarities with simulations of \cite{sujovolsky2018} but they observed a steeper decay $1/f^4$ (at smaller $Re_b$). Oceanic observations \cite{Asaro2000} show rather $1/f^2$ but the forcing of the flow is extremely complex.  One open question at this stage is: would the spectrum exhibit a $1/f^2$ scaling at larger $Re_b$ and at the smallest inertial scales ? A return to isotropy of the velocity spectrum observed at small scales is expected in the SNLST framework if $Re_b$ is large enough \cite{brethouwer2007}.

\begin{figure}[!htb]
\includegraphics[width=8.5cm]{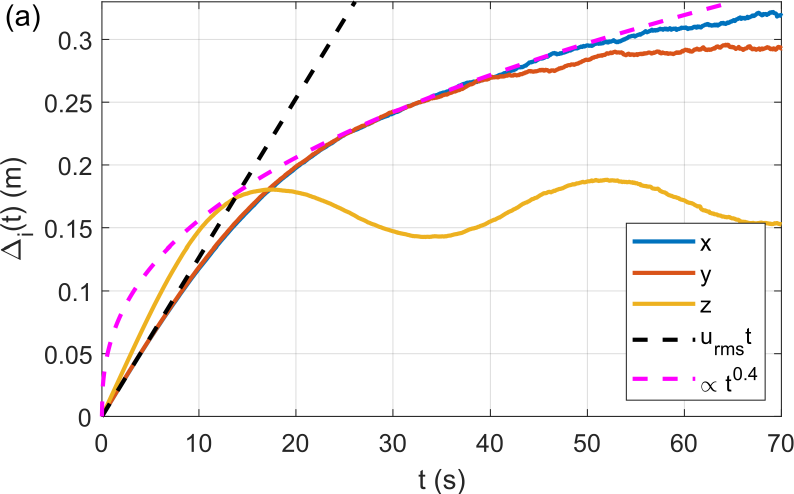}
\includegraphics[width=8cm]{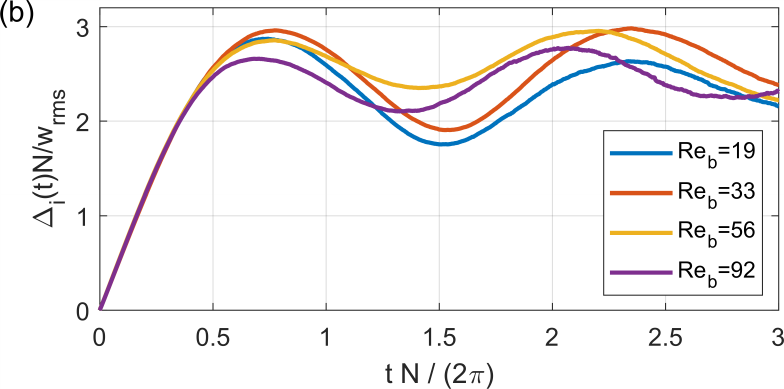}
\includegraphics[width=8cm]{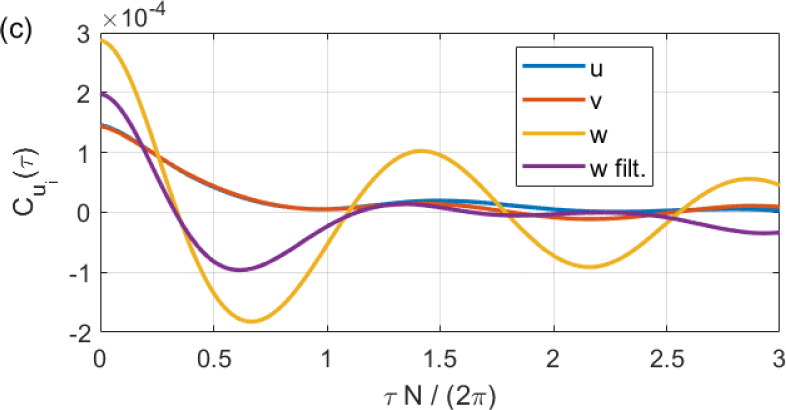}
\caption{\label{disp}Dispersion of single particles. (a) EXP4. The black dashed line is the ballistic regime for the $x$ component $\Delta_u(t)=u_{std}t$. The magenta dashed line is an empirical fit of the data in the diffusive regime with an exponent $0.4$.(b) Rescaled vertical dispersion for the four experiments. (c) Autocovariance of the velocity components for EXP4. The purple curve is the covariance of the vertical velocity after applying a notch filter in the frequency band $[0.67,0.73]N$ (forcing frequencies).}
\end{figure}
The statistics of one particle dispersion are shown in fig.~\ref{disp}. $\Delta_i(t)$ is shown in (a) in the case of EXP4 and for all three components of the position. The difference between vertical and horizontal dispersion is striking. All curves start with a ballistic part with different slopes due to slightly different values of ${u_i}_{std}$ (numerical values are given in SM \cite{SM}). For $t>10$~s the ballistic regime is followed by a short diffusive range and some saturation at long time for the horizontal diffusion. By contrast the vertical diffusion saturates to a plateau with some oscillations around a mean value. The frequency of the oscillation corresponds to $0.7N$ and is a reminiscence of the forcing. The long time saturation of the horizontal dispersion is an experimental artifact due to the finite size of the measurement region: Only particles that do not disperse too far can be measured at long times and thus it induces a long time bias in favor of the particles that remain longer in view of the cameras \cite{Crawford2004}. The vertical dispersion for the four experiments is shown in fig.~\ref{disp}(b) when normalizing the time by $2\pi/N$ and the dispersion by the buoyancy scale $L_b=N/ w_{std}$. One observes a fair collapse of the curves let alone for the forcing oscillations with an asymptotic value of the saturation close to $2.5 L_b$. 

Following Taylor, the long time behavior of the dispersion must be $\Delta_i^2(t)=2{u_i}_{std}^2T^i_L t$. In order to observe a diffusive regime, one must have $T^i_L>0$ but  to observe an asymptotic plateau, one must have $T^i_L=0$. The correlations functions $C_i(\tau)=\langle u_i(t+\tau)u_i(t)\rangle$ are shown in fig.~\ref{disp}(c). The horizontal correlations remain positive and decay to zero similarly to an exponential decay and thus $T^u_L$ and $T^v_L$ are strictly positive as is the case for homogeneous isotropic turbulence. By contrast the vertical correlation becomes negative at $\tau\approx 2\pi/3N$. It shows oscillations at longer time that are again the reminiscence of the forcing frequencies. The purple curve corresponds to $C_w(\tau)$ when filtered by a notch filter removing the forcing frequencies in the interval $[0.67,0.73]N$. The oscillations are no longer visible and the correlation becomes negative and ultimately goes back to zero without oscillations. The numerical integration gives a value of $T^w_L$ which is one order of magnitude lower than the horizontal times, consistently with the observation of the plateau.

\begin{figure}[!htb]
\includegraphics[width=8.5cm]{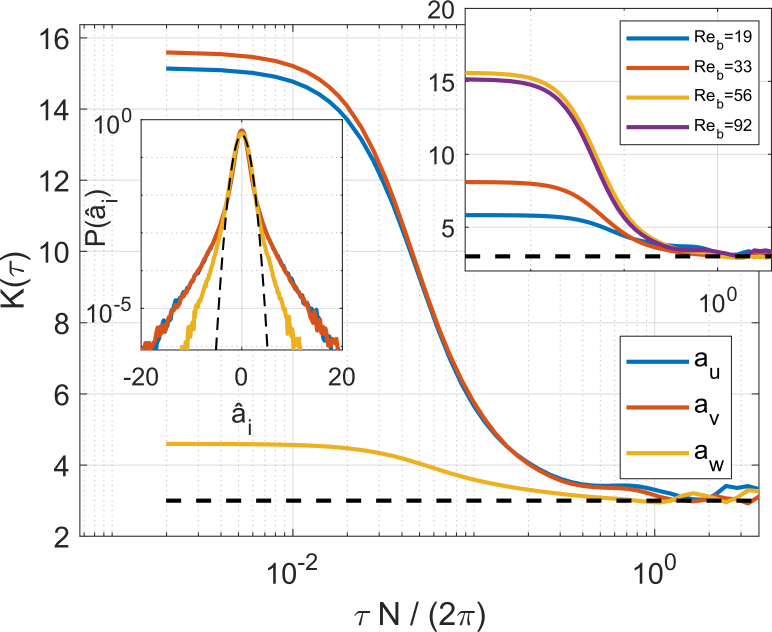}
\caption{\label{inter}
Main figure: evolution of the kurtosis $K$ of the velocity increments as a function of the scale $\tau$ for EXP4. The horizontal dashed line highlights the Gaussian value $K=3$. Right insert: Evolution of the kurtosis of the horizontal increments $\delta_\tau u$ as a function of $\tau$ and $Re_b$. Left insert: PDF of the normalized acceleration components $\hat a_i=a_i/{a_i}_{std}$ for EXP4 (numerical values of ${a_i}_{std}$ are given in SM \cite{SM}). The color for the components are similar to that of the main figure. The black dashed line is a Gaussian distribution.}
\end{figure}

As shown in earlier work, our flow is made of weakly non linear wave turbulence at large scales and strongly nonlinear turbulence at small scales \cite{savaro2020generation,Rodda2022,Rodda2024}. The former is expected to remain close to Gaussian statistics while strong turbulence is characterized by strong intermittency of the Lagrangian statistics \cite{yeung1989,laporta2001,mordant2001}. The usual tool to study intermittency is to compute the statistics of velocity increments $\delta_\tau u_i(t)=u_i(t+\tau)-u_i(\tau)$. The distribution of velocity increments over a large interval of time scales is shown in SM \cite{SM}. The distribution at large scale are Gaussian and wide tails are developed at small scales.  
When $\tau$ goes to zero, the increments become proportional to the Lagrangian acceleration. The distribution of the acceleration is shown in fig.~\ref{inter} (left insert) for EXP4. A wide tail distribution can be observed with significant anisotropy between horizontal and vertical directions. In the SNLST phenomenology, one expects to recover isotropy at the smallest scales if the buoyancy Reynolds number becomes very large. The small scale anisotropy that we observe may be related to a finite $Re_b$ effect and a full isotropy could possibly be observed at larger $Re_b$ (which may not be possible in practice with the current setup, while keeping $F_h$ small enough).

The development of intermittency can be quantified by computing the moments of the distribution of the velocity increments, in particular the 4th order moment. The evolution of the kurtosis $K_i(\tau)=\dfrac{\langle \delta_\tau u_i^4\rangle}{\langle \delta_\tau u_i^2\rangle^2}$ is shown in fig.~\ref{inter}. For $\tau>\pi/N$, the kurtosis is close to the Gaussian value $K_i=3$. When $\tau$ is reduced below $\pi/N$, $K_i$ increases until it reaches a plateau value for $\tau<2\pi/N/100$. The small scale value of the plateau is $K_i\approx17$ for the horizontal components and significantly smaller $K_w\approx 5$ for the vertical component. 
The plateau is reached when $\tau$ lies in the viscous range of scales. The right insert of fig.~\ref{inter} shows the evolution of the kurtosis of $\delta_\tau u$ with the buoyancy Reynolds number. We see a gradual increase of the plateau value except for the highest Reynolds number ($Re_b=92$) that remains slightly below the curve at $Re_b=56$. This could be due to a slight restriction of the resolution at small scales, at which the signal over noise ratio deteriorates and prevents a proper evaluation of the velocity increments at the smallest scales. 

To conclude, we have carried out Lagrangian measurements of stratified turbulence in a large-scale facility, at both high buoyancy Reynolds number and low Froude number, a range of values relevant for turbulent regimes of many geophysical or astrophysical stratfied flows. The saturation of vertical particle dispersion is constrained by the limited amount of kinetic energy available for conversion into potential energy. The saturation level is of order $\Delta_z \approx 2.5 w_{std}/N$, in agreement with previous numerical simulations~\cite{kimura1996,kaneda2000,cambon2004,liechtenstein2006,venayagamoorthy2006,aartrijk2008,brethouwer2009,sujovolsky2018,buaria2020,petropoulos2025}. The prefactor may, however, depend on the range of $Re_b$ and $F_h$ considered. It is also worth noting that our particles are solid and therefore cannot modify their density through molecular diffusion. As a consequence, the ultimate diffusive regime reported, for instance, by \cite{aartrijk2008,brethouwer2009}, cannot be observed here -- even if trajectories were long enough.
The saturation of vertical dispersion has been explained by linear models, such as rapid distortion theory and kinematic simulations, which consider only a superposition of linear gravity waves~\cite{nicolleau2000,cambon2004}. These models attribute the saturation to phase mixing of the modes, associated with the dispersive nature of internal waves. Thus the saturation of dispersion is due to the large scale weak turbulence. Additional support comes from other approaches, including Langevin-type stochastic models that incorporate buoyancy effects~\cite{csanady1964,pearson1983,sujovolsky2018,petropoulos2025}.
The velocity power spectral density is observed to decay as $1/f^3$. This decay is steeper than the $1/f^2$ scaling characteristic of isotropic turbulence, but further investigation is required to establish whether it is an intrinsic effect of buoyancy or a consequence of finite $Re_b$. In the scale range corresponding to wave periods ($\tau > 2\pi/N$), no intermittency is detected, consistent with the emergence of weakly nonlinear wave turbulence at large scales~\cite{Rodda2022}. At smaller scales, however, a non-Gaussian behavior develops, reflecting the presence of strongly nonlinear turbulence driven by wave breaking~\cite{Rodda2024}. While the velocity spectrum becomes isotropic at small scales, intermittency remains anisotropic: the kurtosis of vertical acceleration is lower than that of the horizontal component. A similar anisotropy has been reported in DNS~\cite{buaria2020}, leaving open the question of whether it is a robust feature of stratified turbulence or merely a finite-$Re_b$ effect.

A major issue concerning climate change modelling is to accurately account for the transport and mixing of heat, CO$_2$, or other chemicals into the ocean, which is a major sink for these quantities. We confirm that stratification results in a strongly anisotropic turbulent transport. The horizontal transport remains diffusive and thus may still be modelled by a classical turbulent diffusivity, possibly with some modified numerical prefactors due to the different nature of turbulence. Our experimental findings of non-diffusive vertical transport support numerical results showing that vertical motions in stratified flows are controlled by buoyancy scale dynamics and sensitive to vertical resolution and mixing schemes \cite{waite_2016,cullen_2017}, which highlights the need for anisotropic parametrisation in high-resolution models.

\begin{acknowledgments}
This project received initial financial support from the European Research Council (ERC) under the European Union’s Horizon 2020 research and innovation programme (Grant Agreement No. 647018 – WATU). It is supported by the Simons Foundation through the Simons Collaboration on Wave Turbulence.
We gratefully acknowledge Mickaël Bourgoin and Hugo Pradel for their assistance with the PTV software, and Joël Sommeria for his support with camera calibration.

\end{acknowledgments}

\bibliography{biblio3}

\end{document}